\documentstyle[amstex]{mn}
\newcommand{\apj}{Astrophys. J.}

\input epsf
\def\plotone#1{\centering \leavevmode
\epsfxsize=\columnwidth \epsfbox{#1}}

\def\plottwow#1#2{\centering \leavevmode
\epsfxsize=.49\columnwidth \epsfbox{#1}\epsfxsize=.49\columnwidth \epsfbox{#2}}
\def\plotthree#1#2#3{\centering \leavevmode
\epsfxsize=.99\columnwidth \epsfbox{#1} \hfil
\epsfxsize=.99\columnwidth \epsfbox{#2} \hfil
\epsfxsize=.99\columnwidth \epsfbox{#3}}

\title{3D simulations of shear instabilities in magnetized flows}
\author[]
       {M. Br\"uggen$^{1,2}$, W. Hillebrandt$^{1}$ \\
        $^1$ Max-Planck-Institut f\"ur Astrophysik,
Karl-Schwarzschild-Str.1, 85740 Garching, Germany\\ 
$^2$ Churchill College, Storey's Way, Cambridge, CB3 0DS, UK}

\pagerange{\pageref{firstpage}--\pageref{lastpage}}
\pubyear{1999}

\begin{document}

\maketitle

\begin{abstract}

We present results of three-dimensional (3D) simulations of the
magnetohydrodynamic Kelvin-Helmholtz instability in a stratified shear
layer. The magnetic field is taken to be uniform and parallel to the
shear flow. We describe the evolution of the fluid flow and the
magnetic field for a range of initial conditions. In particular, we
investigate how the mixing rate of the fluid depends on the Richardson
number and the magnetic field strength. It was found that the magnetic
field can enhance as well as suppress mixing. Moreover, we have
performed two-dimensional (2D) simulations and discuss some interesting
differences between the 2D and 3D results.

\end{abstract}



\section{Introduction}

Observational evidence seems to suggest that current formalisms
underestimate the efficiency of the mixing processes that operate in
stars, especially in fast rotating stars. Therewhile, the
observational evidence for mixing is increasing rapidly. Herrero et
al. (1992) find that all fast rotating O-stars show significant
surface He-enrichments.  Other observations include the N/C and
$^{13}$C and $^{12}$C enrichments of stars on the Red Giant Branch
(e.g. Kraft et al. 1997, Charbonnel 1995), the He and N excesses in OBA
supergiants (Fransson et al. 1989), the depletion of boron in most
B-type stars (Venn, Lambert \& Lemke 1996) and the ratio of the number
of blue to red supergiants in galaxies (Langer \& Maeder 1995). For a
more detailed review see Maeder (1995), Kraft (1994) and references
therein. These observations seem to suggest that mixing is strong
enough to transport the nuclearly processed material to the surface in
a fraction of the life time of the star. It has been shown that, both,
the enrichment in CNO elements and the depletion of fragile elements
such as boron can be explained if some form of mixing is introduced
(Langer 1992, Denissenkov 1994, Denissenkov \& Tout 2000, Meynet \& Maeder
1997).\\

The Kelvin-Helmholtz instability occurs in chemically homogeneous,
stratified shear flows when the destabilising effect of the relative
motion in the different layers dominates over the stabilising effect
of buoyancy (see e.g. Chandrasekhar 1961).  The competition between
the two effects is described by the Richardson number, Ri:

\begin{equation}
{\rm Ri} = \frac{g\rho'}{\rho {U'}^2},
\end{equation}
where $\rho$ is density, $U$ the shear velocity and $g$ the
gravitational acceleration. The prime indicates the derivative in the
direction of gravity. Maeder (1995, 1997) and Maeder \& Zahn (1998)
generalised the Richardson criterion to include radiative losses and
changes in the chemical potential. For a further discussion of the
Richardson number the reader is referred to their papers.\\

Magnetic fields in stars are believed to be produced by differential
rotation and are predominantly toroidal (e.g. Parker
1979). In the Sun it is thought that the magnetic field is
concentrated at the bottom of the convection zone. The bottom of the
convection zone is also the perceived location of the solar dynamo
which is believed to be responsible for the solar cycle. Within the
convection zone flux expulsion and magnetic buoyancy swiftly remove
any toroidal magnetic field.\\

Seismic measurements of the rotation rate of the Sun have revealed a
differentially rotating convection zone and a rigidly rotating
interior with a shear layer that separates the two. However, the
explanations for the peculiar rotation rates of the Sun remain
controversial.  The most convincing proposals involve shear-generated
quasi-horizontal turbulence (Spiegel \& Zahn 1992) or a large-scale
magnetic field in the radiative interior (Gough \& McIntyre 1998). The
differential rotation of the Sun and other stars is only one of the
issues that demonstrate the importance of simulating shear
instabilities in the presence of magnetic fields.  The magnetic
Kelvin-Helmholtz instability has been treated in other physical
contexts, for example in the heliosphere, where the solar wind flows
past planetary magnetospheres (see, e.g., Uberoi 1984), in the context
of the stability of interstellar clouds (Vietri, Ferrara \& Miniati
1997) and in accretion disks (Anzer \& B\"orner 1983).\\

Apart from mixing by shear instabilities, there is another important
candidate believed to be reponsible for extra mixing in stars, namely
convective overshoot.  This mechanism has been investigated
numerically by a number of groups (Freytag, Ludwig \& Steffen 1996,
Nordlund \& Stein 1996, Singh, Roxburgh \& Chan 1998) all of which
find some degree of overshoot.  Observations seem to support this:
Isochrone fitting to stellar clusters and binary systems suggests that
convective overshoot is significant (see Zahn 1991 for a review).\\

In this paper we simulate shear instabilities in a stratified
magnetized fluid.  The magnetic field can change the picture that was
obtained in the unmagnetized case (see Br\"uggen \& Hillebrandt 2000)
in various ways: The instability can locally enhance or diminish the
magnetic field, which, in turn, can enhance dissipation by
reconnection and alter the flow pattern.

When the field is parallel to the shear flow, magnetic tension will
have a stabilising influence on the flow.  This can be regarded as a
consequence of the field line tension whose effect is similar to the
effect of surface tension, e.g. in Rayleigh-Taylor instabilities.  If
the field is perpendicular to the direction of the flow, it affects
the flow only through an extra pressure which changes the magnetosonic
speed.  A linear analysis of the magnetohydrodynamic shear instability
was presented by Chandrasekhar (1961).\\

In the incompressible case it can be shown that a uniform magnetic
field, that is aligned with the shear flow, stabilises the
Kelvin-Helmholtz instability as long as the velocity jump across the
shear layer is less than twice the Alfv\'en speed.  For the
compressible case, Miura \& Pritchett (1982) presented a linear
stability analysis, but in order to proceed further, numerical
simulations were required.

Malagoli et al. (1996) performed a set of 2D MHD simulations of the KH
instability in a uniform magnetic field where they varied the ratio of
the Alfv\'en to the sound speed.  They could identify three stages in
which the instability develops: the linear stage, the dissipative
transient stage with intermittent reconnection events and the
saturation stage, where the turbulence decays into aligned structures.
These findings were confirmed by Keppens et al. (1999). These authors
reported 2D MHD simulations for the case where the magnetic field was
unidirectional everywhere and for the case where the magnetic field
changes sign in the middle of the shear layer. The two cases were
found to be dynamically very different.\\

In a series of papers Jeong et al. (2000), Jones et al. (1997) and
Frank et al. (1996) presented 2- and 2.5-dimensional simulations of
uniform shear layers in the presence of magnetic fields of varying
strength and direction.  In the non-magnetic case it is known that
shear flows develop a vortex, also known as ``Cat's Eye'', which would
spin for a long time before viscous dissipation would eventually
dissolve it.  In the MHD case, however, the authors cited above find
that the quasi-steady state of the flow is a nearly laminar layer
instead of a single big vortex. The magnetic tension stabilises the
flow before the vortex can form. Instead, a broad shear layer develops.
This may not be too surprising for strong magnetic field but it was
not obvious that even very weak fields can fundamentally alter the
flow. For fields just below the critical field strength the
aforementioned authors found that enhancements of the magnetic field
through linear growth can stabilise the flow before it becomes
nonlinear. Even fields that are by a factor of 2.5 weaker than the
critical field strength were found to reconnect and reorganise the
flow such as to quickly relax into a marginally stable laminar flow.
Even though the magnetic field was initially weak, it was strong
enough for magnetic stresses to become important before the formation
of the primary vortex.\\

Very recently, Ryu, Jones \& Frank (2000) published results of 3D
simulations of the nonlinear evolution of of a uniform (unstratified)
shear layer. They performed high resolution simulations mainly with
the intention to study the turbulent properties of uniform MHD
flows. We, on the other hand, did not aim to simulate full MHD
turbulence, for which our resolution would be insufficient. Instead,
we make a first attempt at quantifying mixing rates in magnetized,
compressible and stratified shear flows as a function of the
Richardson number. For this purpose we performed a set of 3D and 2D
MHD simulations and investigated the dependence of the mixing rate on
the Richardson number and the magnetic field strength.  Moreover, we
studied the nonlinear dynamics of the magnetic shear instability and
investigated the differences between 2D and 3D simulations.  With
conditions in stellar interiors in mind, we restricted ourselves to
subsonic flows.  As far as we know, to date, these are the first 3D
simulations of the magnetohydrodynamic Kelvin-Helmholtz instability in
a {\it stratified} fluid.

\section{Numerical simulations}

Here we present the results of numerical simulations of the
magnetohydrodynamic shear instability in a stratified fluid.  The
simulations were obtained using the ZEUS-3D code which was developed
especially for problems in astrophysical hydrodynamics (Clarke \&
Norman 1994).  The code uses finite differencing on a Eulerian or
pseudo-Lagrangian grid and is fully explicit in time. It is based on
an operator-split scheme with piecewise linear functions for the
fundamental variables. The fluid is advected through a mesh using the
upwind, monotonic interpolation scheme of van Leer.  The magnetic
field is evolved using a modified constrained transport technique
which ensures that the field remains divergence-free to machine
precision. The electromotive forces are computed via upwind
differencing along Alfv\'en characteristics. For a detailed
description of the algorithms and their numerical implementation see
Stone \& Norman (1992a, b).\\

\begin{figure}
\plotone{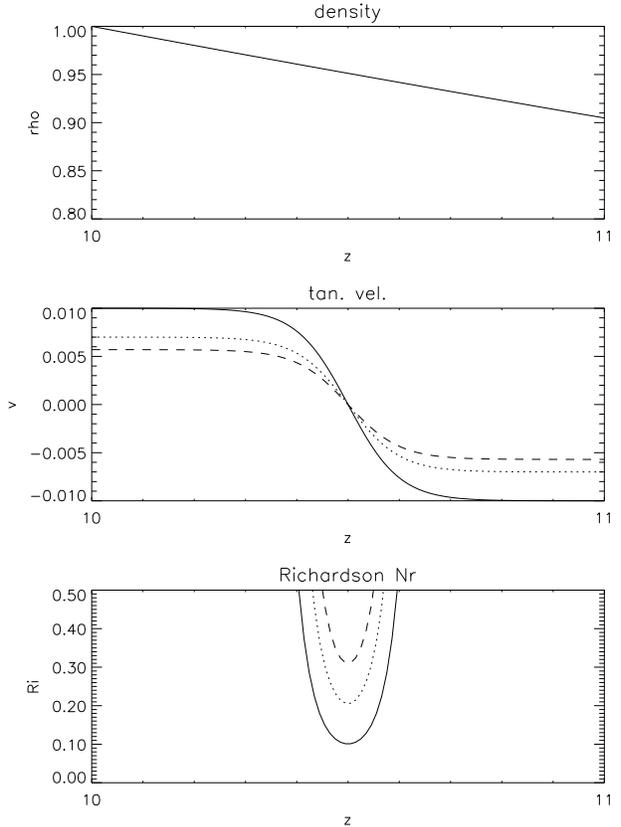}
\caption{Initial density and shear velocity distributions. The bottom
panel shows the Richardson numbers for the respective velocities
(correspondence denoted by the same line style). 
}
\label{ini}
\end{figure}

As initial model, an isothermal density distribution under constant
gravitational acceleration in hydrostatic equilibrium was chosen. The
initial density distribution is shown in Fig. \ref{ini}. In the
following, we will use dimensionless units which are determined by a
gravitational constant of $G=1.$, a gravitational acceleration of
$g=0.01$ and a density scale height of 10. Then a shear velocity
profile was imposed on the fluid. It was assumed to have the form of a
hyperbolic tangens in order to minimise the effect of the boundaries
onto the shear layer, i.e.

\begin{equation}
U(z) = U_0\tanh [(z-z_0)/h] ,
\end{equation}
where $U_0$ is the amplitude of the shear ($x$-) velocity, $z$ the
vertical position of the shear layer, and $h$ its extent. In order to
keep the shear layer away from the boundaries, $h$ was taken to be
smaller than the vertical extent of the simulation region. $U_0$
was chosen to yield a range of initial Richardson numbers of 0.05 -
0.3, where the Richardson number is taken in its original simple
definition, i.e. Ri$=g\rho'/\rho {U'}^2$ and is measured at $z=z_0$.
Finally, $B$ was chosen to yield a range of Alfv\'en velocities which
extend up to the velocity jump across the shear layer (see Table 1 and
2 for details). As shown in Table 2, the magnetic energy ranges from
$\sim 1$ \% of the kinetic energy to a few times the kinetic energy. In
this paper we only consider the case where the initial magnetic field
is uniform and parallel to the shear flow (unidirectional
everywhere). For comparison, we also repeated the simulations with the
magnetic field set to zero. Fig. \ref{ini} shows the initial
conditions for Ri = 0.1, 0.2, and 0.3. The boundary conditions were
chosen to be periodic in the $x-$ direction and reflecting in the $y-$
and $z-$ direction.

In our simulations we employed an ideal gas equation of state, we
ignored the effects of rotation, nuclear reactions and radiative
processes.  The simulations were computed on a Cartesian grid and the
computational domain had the dimensions $2$ $\times$ $1$
$\times$ $1$ (in the $x$-, $y$- and $z$-direction, where gravity acts
in the $z$-direction).  It was covered by 100 $\times$ 50 $\times$ 50
grid points.  For comparison, we also performed 2D simulations on a
grid with 200 $\times$ 100 grid points. The calculations were
performed on a CRAY T3E and an IBM RS/6000 cluster.\\

\begin{table}
 \caption{Initial parameters of the simulations: ratio of the mean
 Alfv\'en velocity, $v_{\rm A}$, to the adiabatic sound speed, $c_{\rm
 s}$, and ratios of the mean Alfv\'en velocities to the velocity jump
 across the shear layer for Ri = 0.1, 0.2 and 0.3.}
 
\begin{tabular}{@{}lcccc} 
& $\frac{v_{\rm A}}{c_{\rm s}}$ & $\frac{v_{\rm A}}{\Delta v}$ (0.1) & $\frac{v_{\rm
A}}{\Delta v}$ (0.2) & $\frac{v_{\rm A}}{\Delta v}$ (0.3)\\ \hline \\ 
$B=0.001$ & $2.6\ 10^{-3}$ & 0.05 & 0.07 & 0.09 \\ \hline \\ 
$B=0.003$ & $7.5\ 10^{-3}$ & 0.15 & 0.21 & 0.27 \\ \hline \\ 
$B=0.01$ & $2.6\ 10^{-2}$ & 0.50 & 0.71 & 0.91 \\
\hline 

\end{tabular}

\end{table}

\begin{table}
 \caption{Initial parameters of the simulations: ratio of the initial magnetic
 energy to the initial kinetic energy for different Richardson numbers.}
 \begin{tabular}{@{}lcccc}
  Ri & 0.1 & 0.15 & 0.2 & 0.3  \\ \hline \\
  $B=0.001$ & $1.32\ 10^{-2}$ & $2.04\ 10^{-2}$ & $2.70\ 10^{-2}$ 
          & $4.01\ 10^{-2}$  \\ \hline \\
  $B=0.003$ & 0.118 & 0.183 & 0.243 & 0.360  \\ \hline \\
  $B=0.01$ & 1.316 & 2.037 & 2.699 & 4.012\\ \hline
 \end{tabular}

\end{table}

In order to study mixing processes, the ZEUS code was modified to
follow the motion of 1000 `tracer' particles. They are advected with
the fluid and are initially located in a plane perpendicular to the
$z$-axis in the centre of the shear layer.

\section{Results and Discussion}

First, we will briefly present the results of our 2D simulations and
discuss how they compare with the previous numerical work that was
mentioned in Sec. 1. Then we will compare the 2D simulations with the
3D simulations. In the non-magnetic case (Br\"uggen \& Hillebrandt
2000), we found that the 2D and 3D simulations yielded very similar
results for the mixing rates. Here we will investigate whether this is
still the case when magnetic fields are involved.

\subsection{2D simulations}

\begin{figure}
\plotone{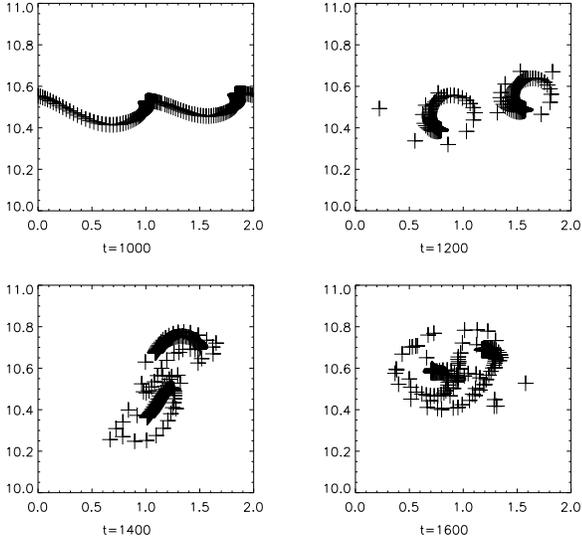}
\caption{Positions of tracer particles at various times for a 2D simulation with
Ri=0.2 and no magnetic field. 
}
\label{tr2D_0}
\end{figure}

\begin{figure}
\plotone{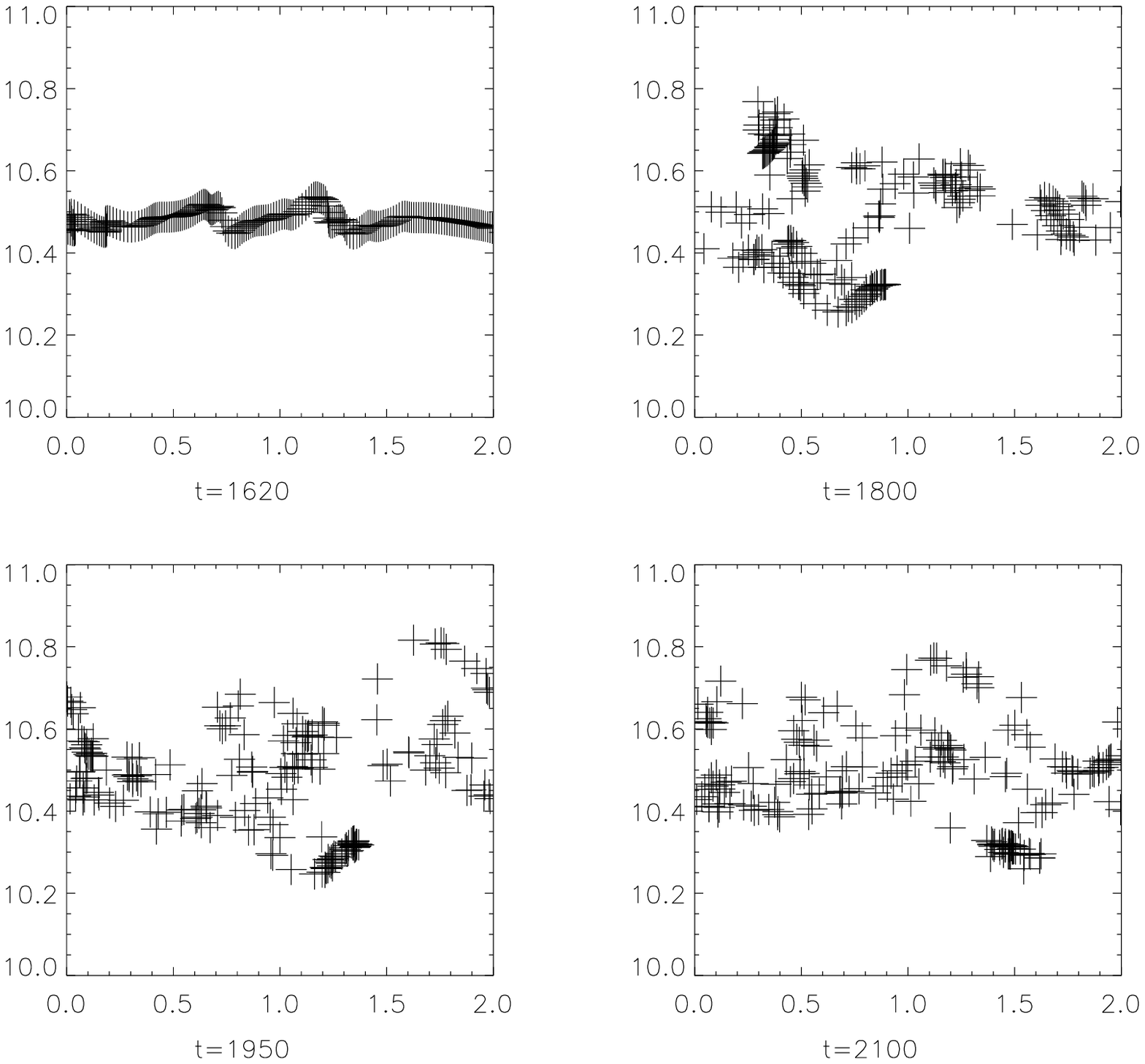}
\caption{Positions of tracer particles at various times for a 2D simulation with
Ri=0.2 and B=0.001. 
}
\label{tr2D_1}
\end{figure}

\begin{figure}
\plotone{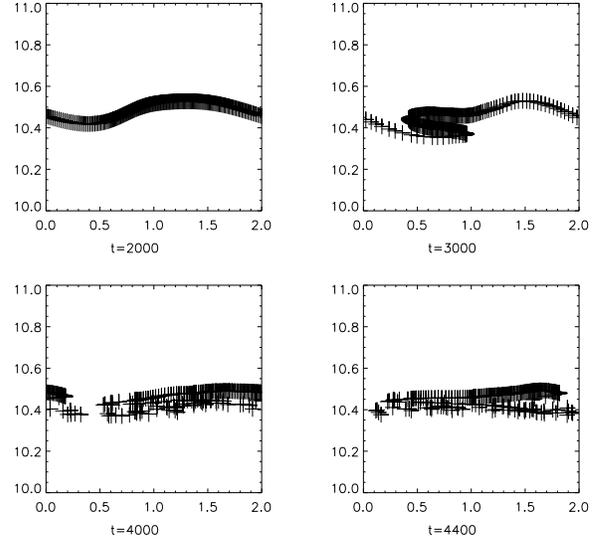}
\caption{Positions of tracer particles at various times for a 2D simulation with
Ri=0.2 and B=0.003. 
}
\label{tr2D_2}
\end{figure}

In general, we found that the effect of the magnetic field on the
shear instability is multifarious and diverse: The magnetic field can
enhance as well as diminish the mixing, depending on the strength of
the shear. Fig. \ref{tr2D_0}-\ref{tr2D_2} show the positions of tracer
particles at various times in some 2D simulations with a Richardson
number of Ri=0.2, for magnetic fields of $B=0$, $B=0.001$ and
$B=0.003$, respectively. Immediately, one can note that the magnetic
field suppresses the formation of the characteristic primary vortex at
the centre of the shear layer and inhibits the mixing. Instead of a
vortex a broad laminar mixing layer develops. The tracer particles
remain within the shear layer and are not swirled around as much as in
the case without a magnetic field. This suppression becomes more
effective the stronger the magnetic fields is. Stronger magnetic
fields lead to a narrower shear layer as the Maxwell stresses tend to
align the magnetic and velocity fields. This is also apparent in the
transverse ($z$-) velocities and the vorticity. The maximum
$z$-velocities and vorticities decrease with increasing magnetic field
strength. A strong magnetic field tends to stabilize any slight
transverse motion before a vortex can form.  Relatively weak field are
being wound up by the shear flow.  During this winding-up the magnetic
field strength locally exceeds its initial value by up to a factor of
$\sim 12$ for the $B=0.001$ and Ri=0.1 run.  For the runs with a
higher Richardson number or higher magnetic fields this enhancement
decreases. We should point out that this amplification of the field is
not the result of some dynamo action. The local enhancement of the
magnetic field comes from the stretching and twisting of the magnetic
field lines. As one may have expected, we find that regions of high
vorticity and enhanced magnetic fields are correlated.\\

\begin{figure}
\plotthree{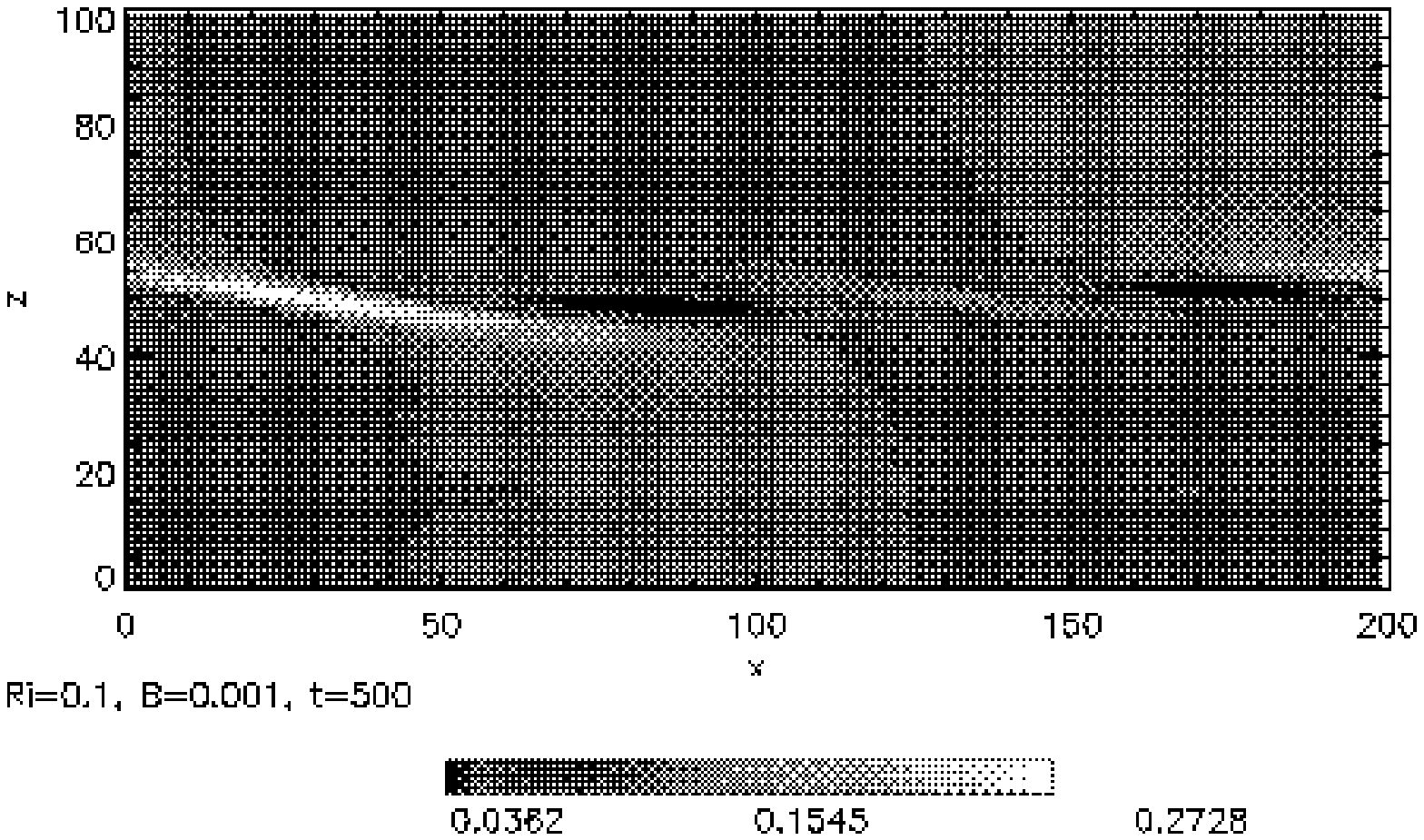}{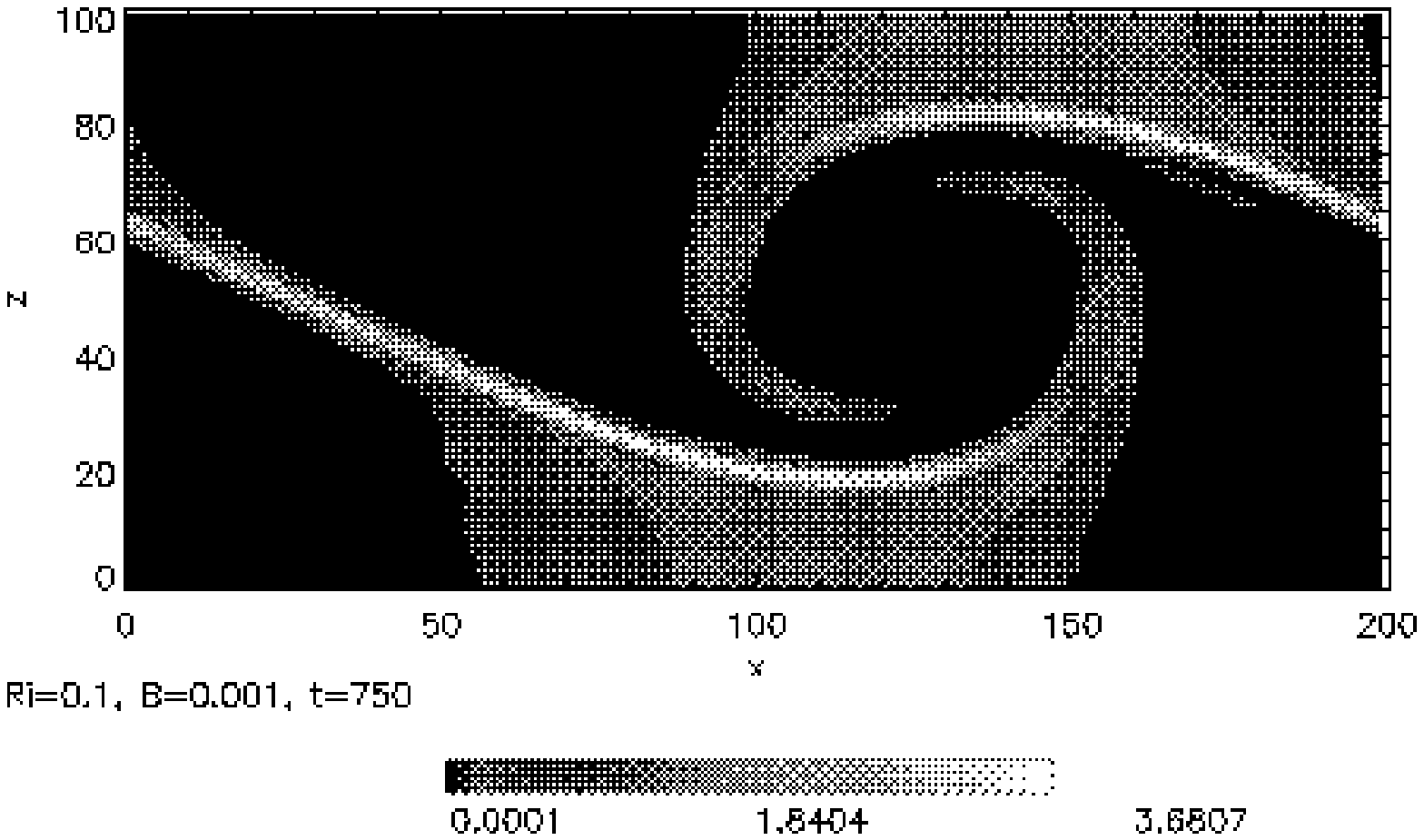}{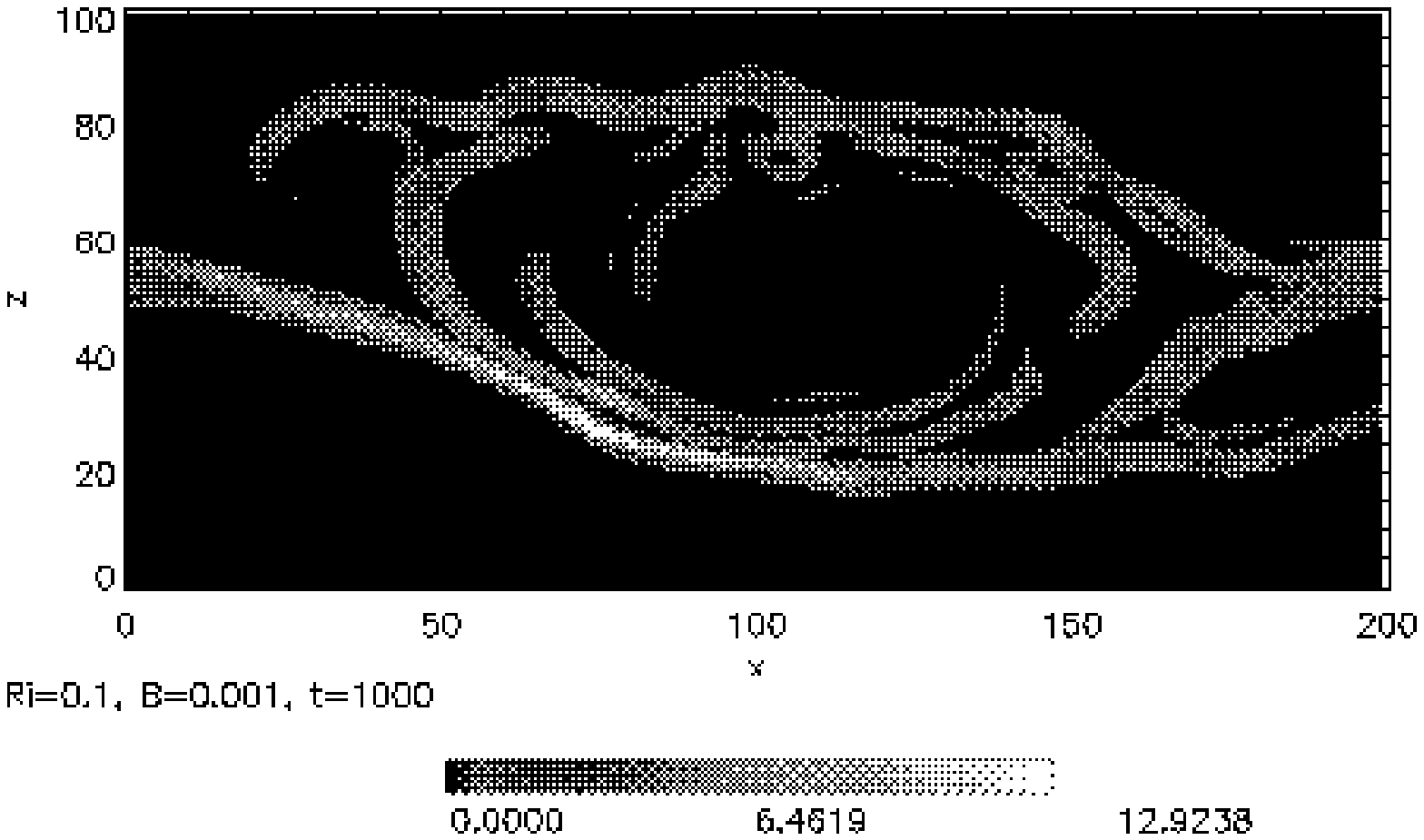}
\caption{Magnetic energy in a 2D simulation with Ri=0.1 and $B=0.001$.
}
\label{me1}
\end{figure}

\begin{figure}
\plotthree{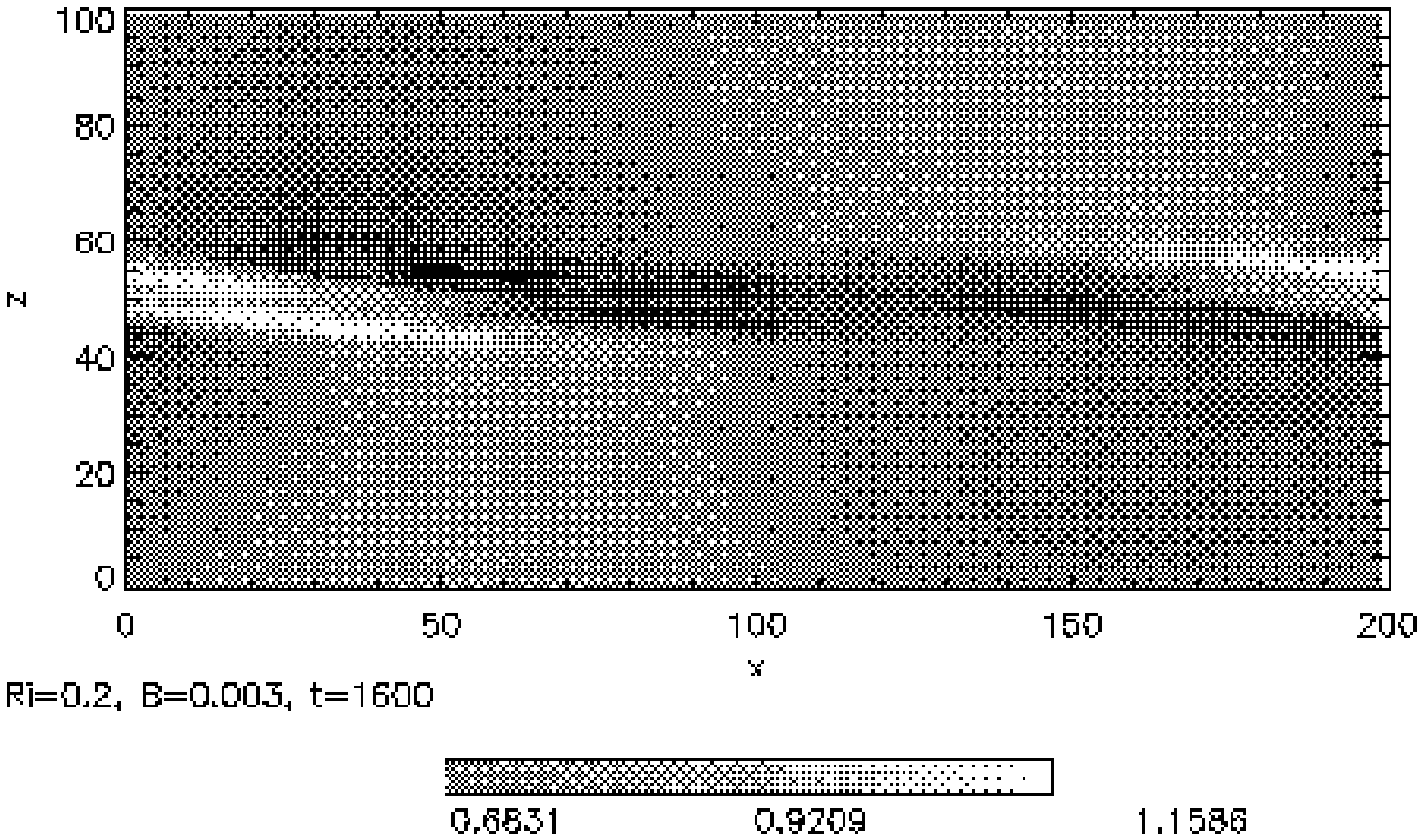}{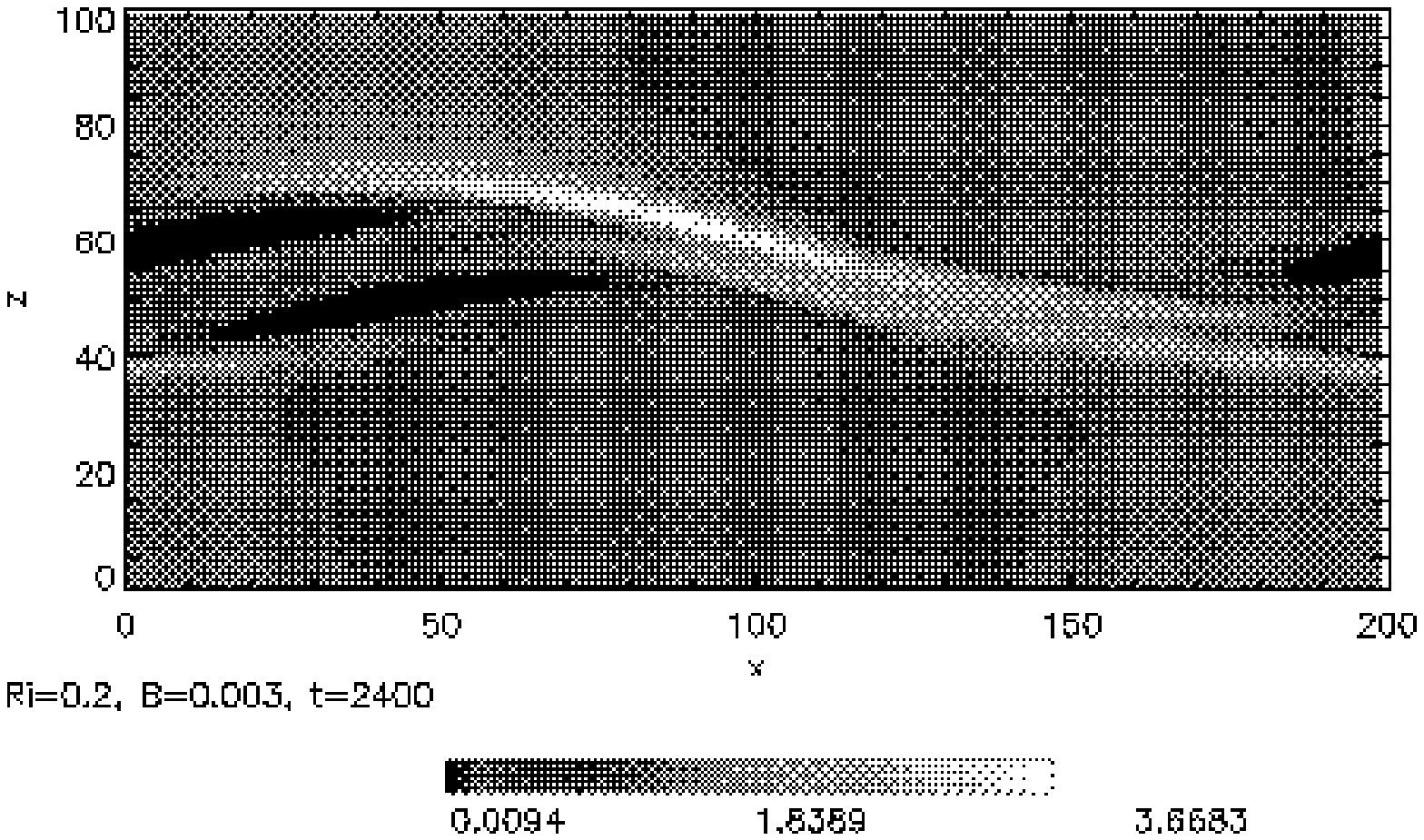}{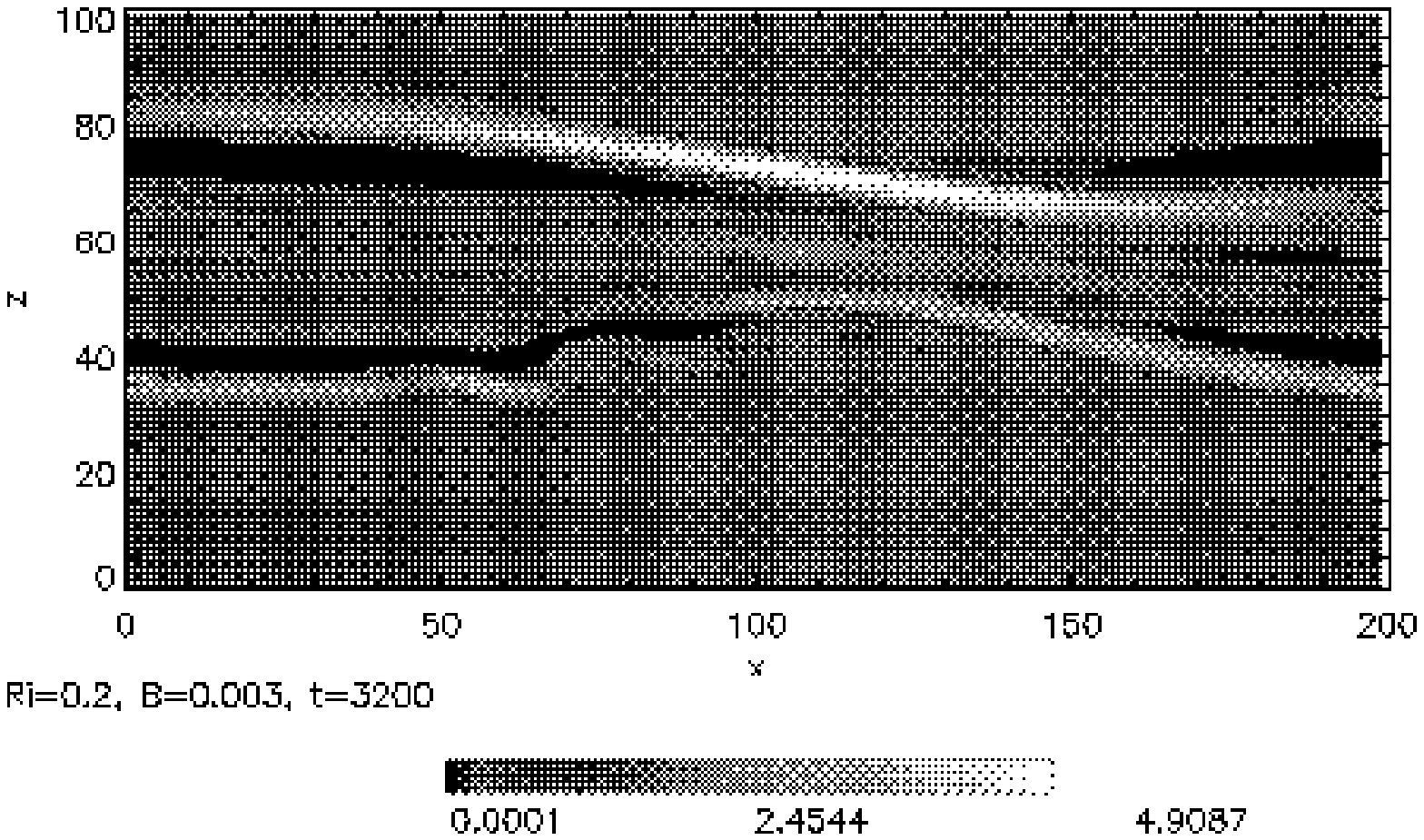}
\caption{Magnetic energy in a 2D simulation with Ri=0.2 and $B=0.003$.
}
\label{me2}
\end{figure}

To visualise the working and evolution of the magnetic field, we show
the magnetic energy for two quite different examples. Fig. \ref{me1}
shows the magnetic energy for a 'weak' field in which the dynamic
effect of the magnetic field is relatively small with Ri=0.1,
$B=0.001$. A different situation is seen in Fig. \ref{me2} which shows
the evolution of a 'strong' field with Ri=0.2, $B=0.003$. In the
latter case, the magnetic field efficiently reduces the mixing in the
shear layer. In Fig. \ref{me1} one can see that a big vortex forms,
which subsequently decays, while in Fig. \ref{me2} the field undulates
a little but remains essentially laminar. In both cases the magnetic
energy increases with time, and this increase is stronger for the more
dynamic case shown in Fig. \ref{me1}. In the weak field case the
magnetic flux is expelled from the centre of the vortex.

For the very strong magnetic field of $B=0.01$, all shear
flows with Ri $\geq$ 1 remained stable for well over 3000 sound
crossing times where we stopped the simulation.\\

This behaviour agrees qualitatively with the works of Frank et
al. (2000) and Jeong et al. (2000) although their initial conditions
somewhat differ from ours. The reader is referred to their paper for
further details of the Kelvin-Helmholtz instability in 2D
simulations.\\

\begin{figure}
\plotone{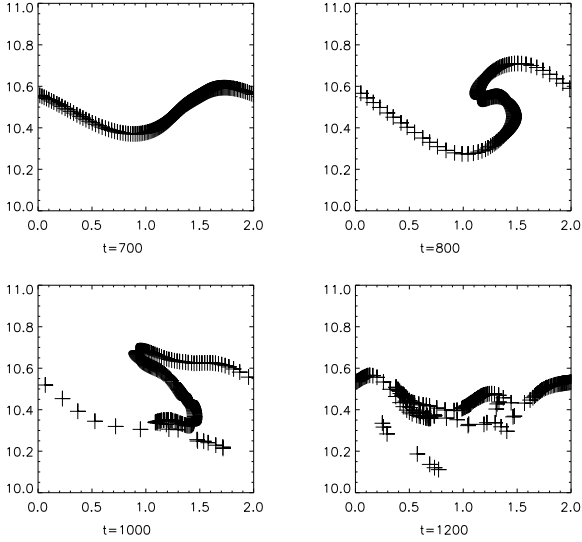}
\caption{Positions of tracer particles at various times for a 2D simulation with
Ri=0.1 and B=0.003. 
}
\label{tr2D_3}
\end{figure}

\begin{figure}
\plotthree{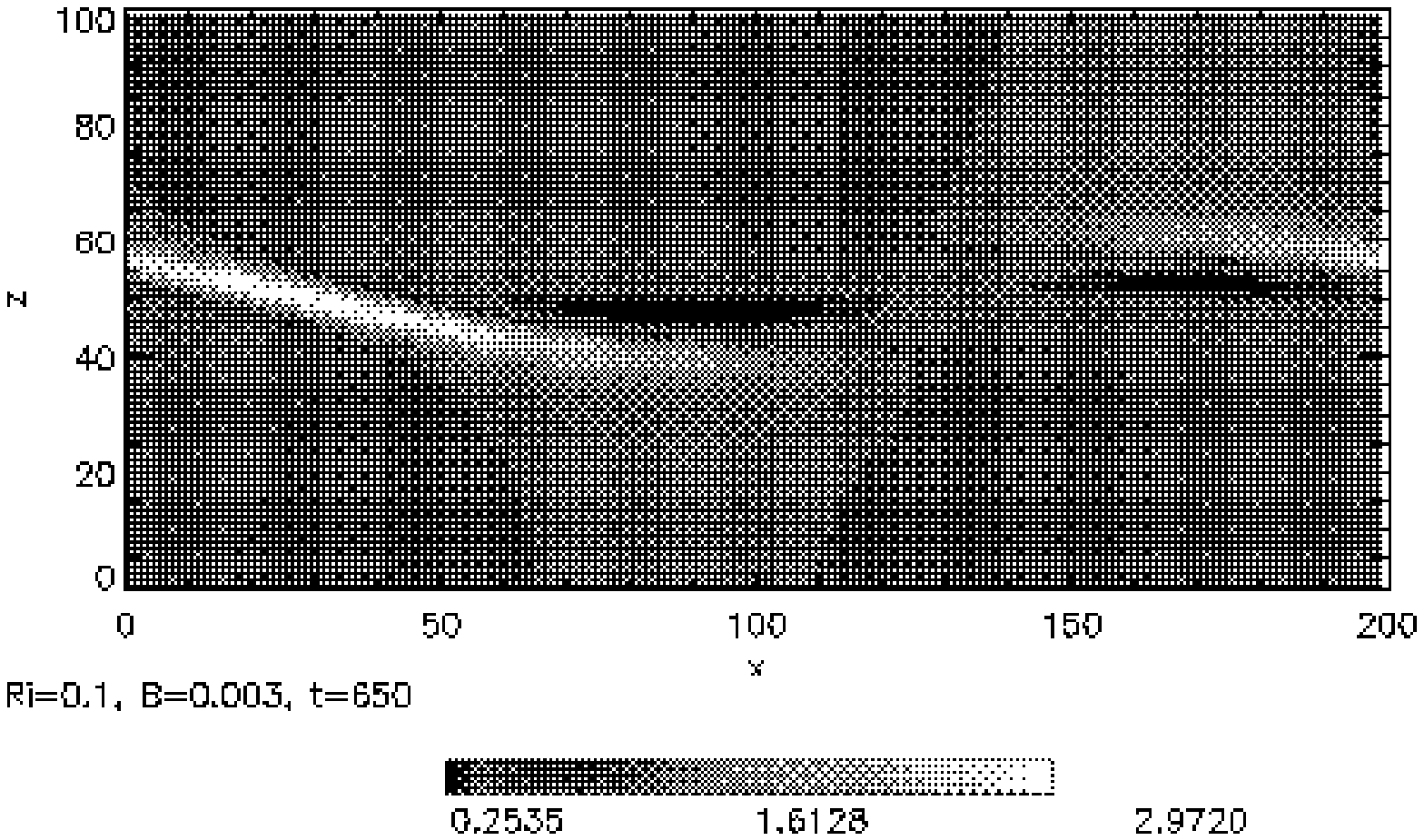}{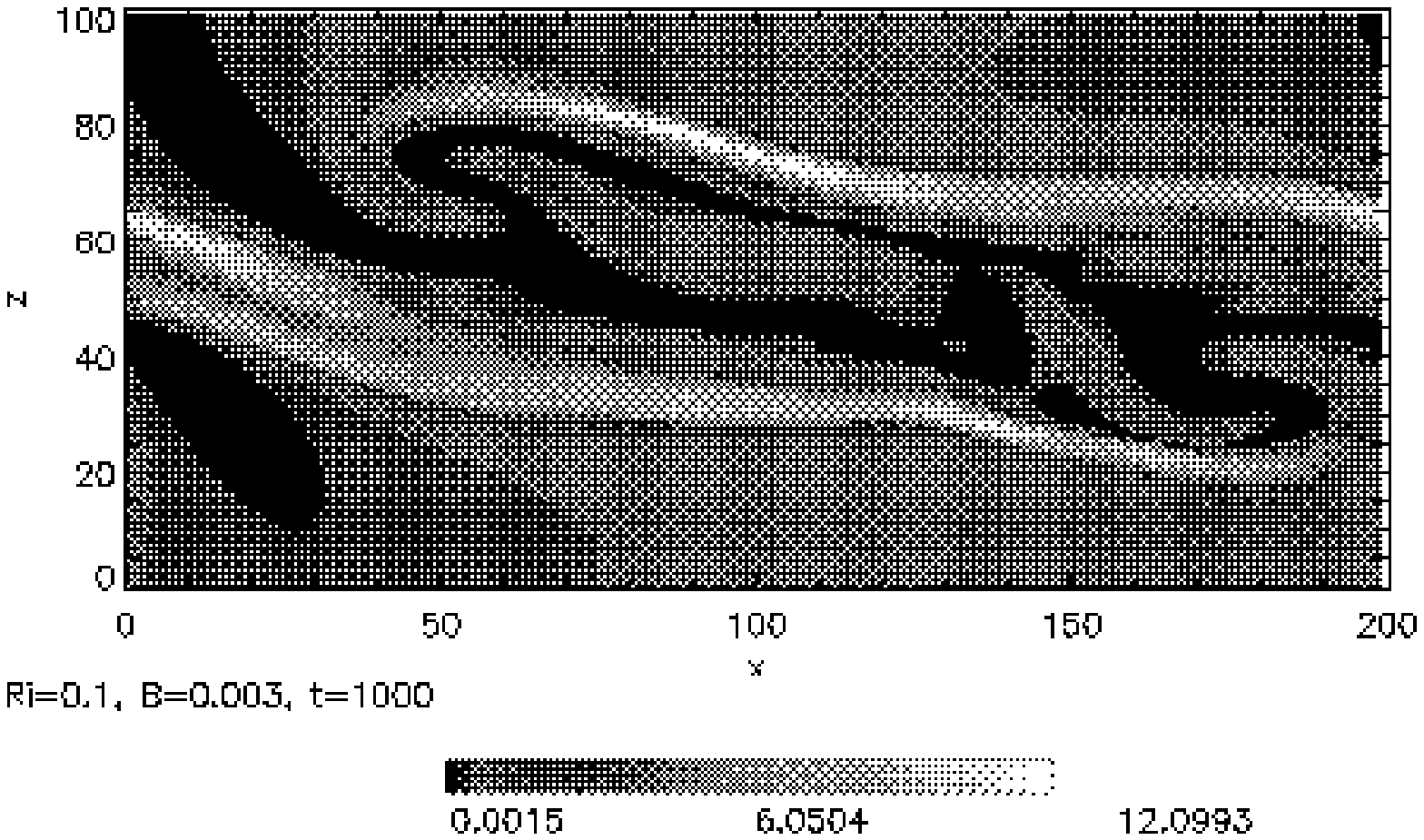}{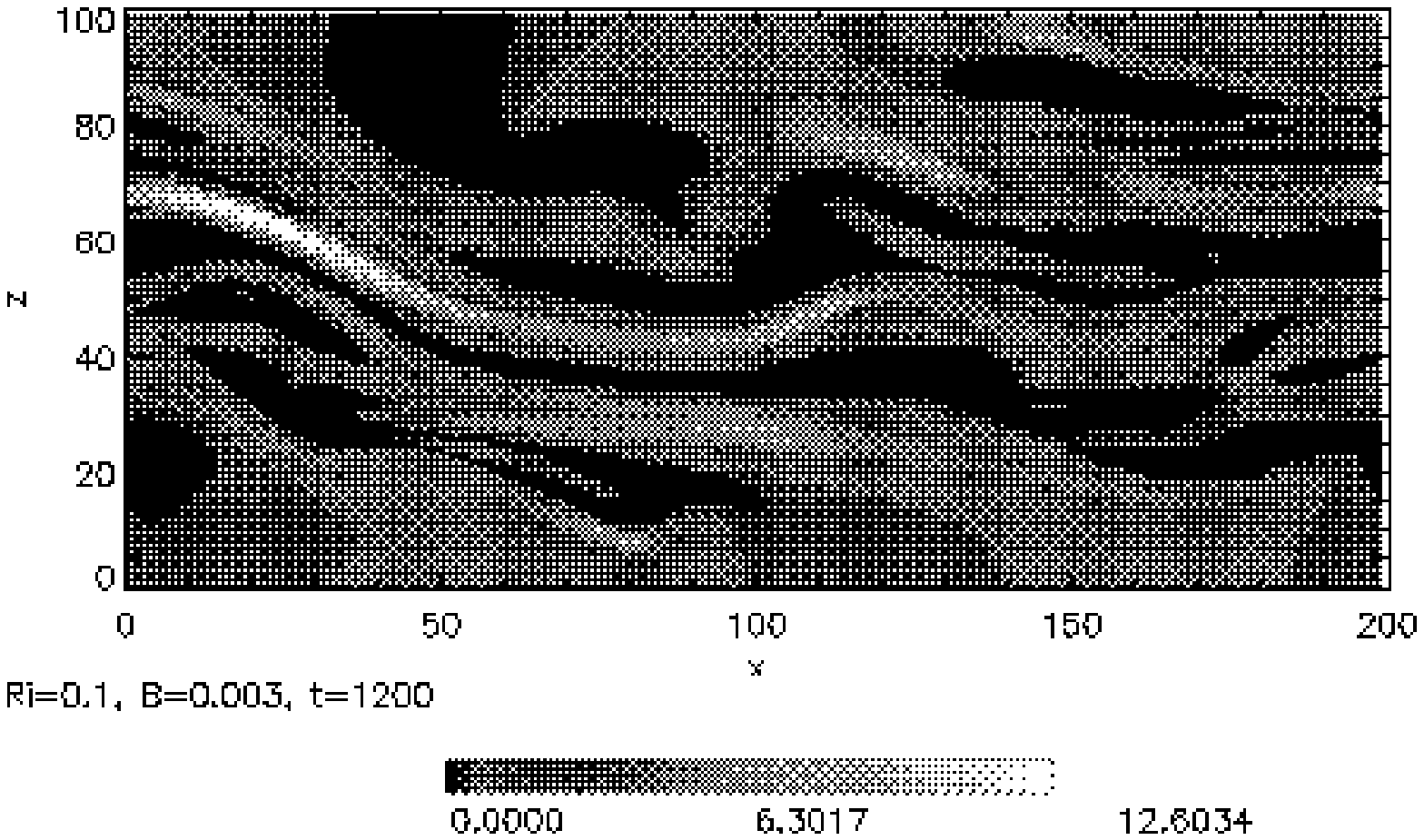}
\caption{Magnetic energy in a 2D simulation with Ri=0.1 and $B=0.003$.
}
\label{me3}
\end{figure}

While in the simulations with Richardson numbers $> 1$ the effect of
the magnetic field was to suppress the mixing, we found the effect to
be reversed for the case with a low Richardson number and a strong
magnetic field (Ri=0.1 and $B=0.003$). Now the magnetic field seems to
enhance the mixing. The positions of the tracer particles are shown in
Fig. \ref{tr2D_3} and the magnetic energy is plotted in
Fig. \ref{me3}. Again, one can note that the magnetic field is bundled
together in two flux tubes which form at the centre of the shear layer
and are aligned with the flow. Subsequently, the flux tubes drift to
the edges of the shear layer and, in doing so, mix the fluid. This can
be seen in the distribution of the tracer particles in
Fig. \ref{tr2D_3}. Eventually, numerical reconnection occurs and the
magnetic field forms a filamentary structure in the shear layer.

\begin{figure}
\plotone{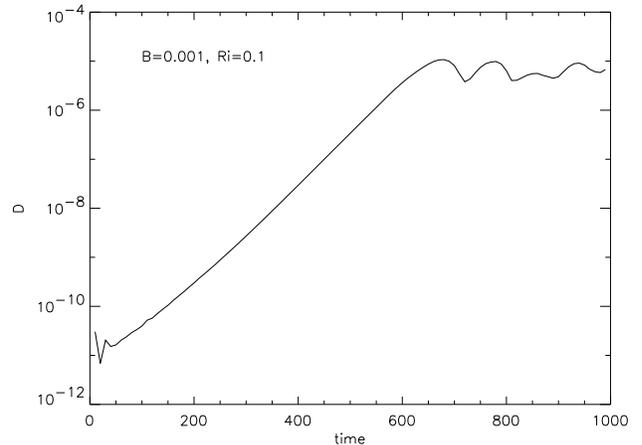}
\caption{Dimensionless diffusion coefficient as a function of time
for Ri=0.2 and B=0.001 (2D simulation). 
}
\label{diff1}
\end{figure}

In Sec. 1, we mentioned the importance of mixing in the context of
stellar evolution. For this purpose, it is useful to quantify the rate
of mixing by some kind of one-dimensional diffusive approximation,
because most stellar evolution codes implement the mixing through a
diffusion equation.  A heuristic diffusion constant can then be
defined as follows (Br\"uggen \& Hillebrandt 2000)

\begin{equation} 
D = \sigma^2/t ,
\end{equation}
where $\sigma^2=\frac{1}{N}\sum_N [z(N)-z_0]^2$, $N$ being the number
of tracer particles, $z_0$ the original height of the tracer particle
and $z$ its height after a time $t$. The diffusion coefficient as a
function of time has been plotted in Fig. \ref{diff1} for one
particular example of parameters. As in the non-magnetic case
described in Br\"uggen \& Hillebrandt (2000), the diffusion
coefficient rises with time before it reaches its maximum. Then it
remains approximately constant, apart from some turbulent scatter,
before it slowly starts to decrease again.  The nearly constant value
to which $D$ is converging is the value which is of greatest interest
for the purpose of evolving stellar models.

\begin{figure}
\plottwow{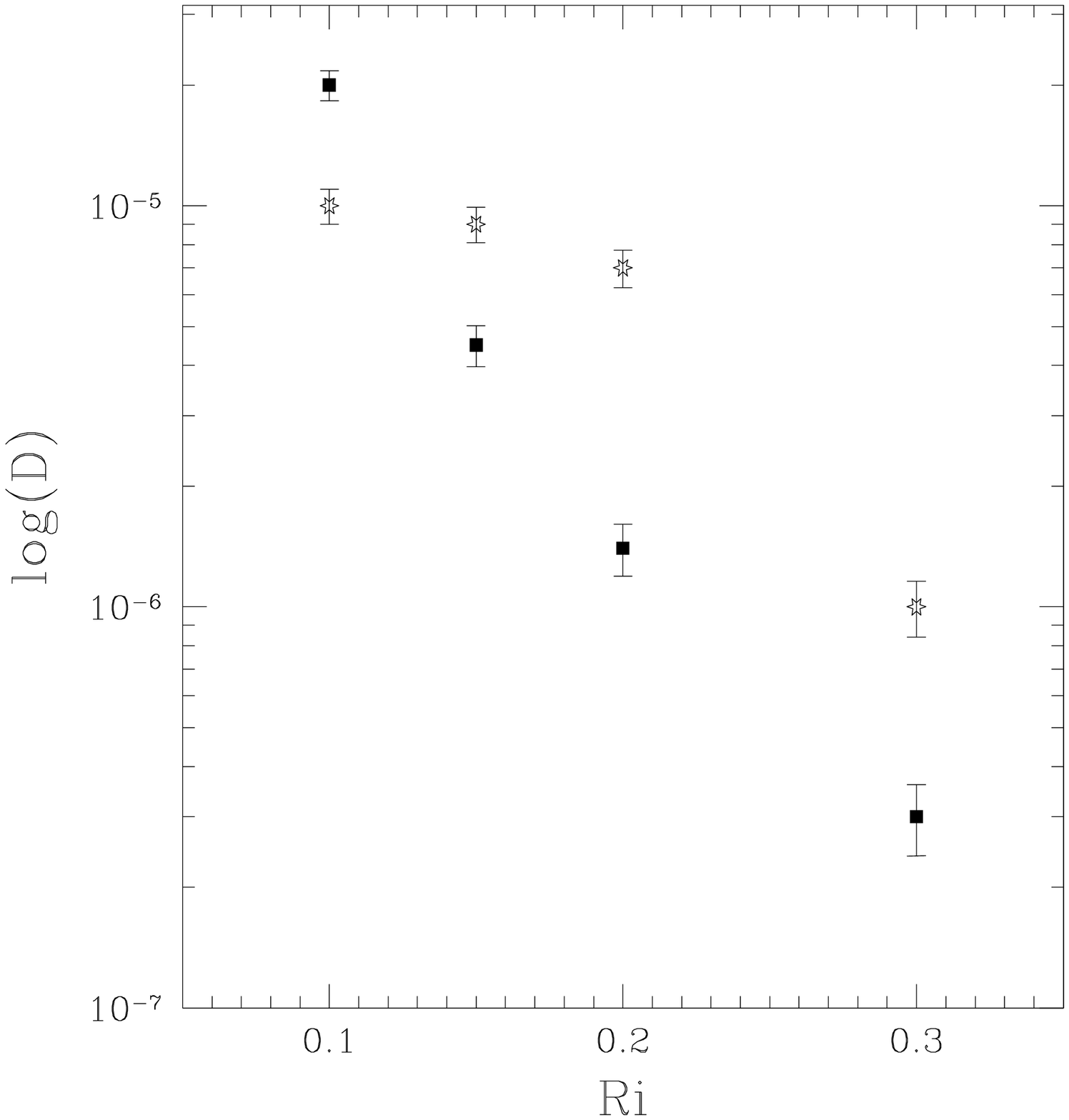}{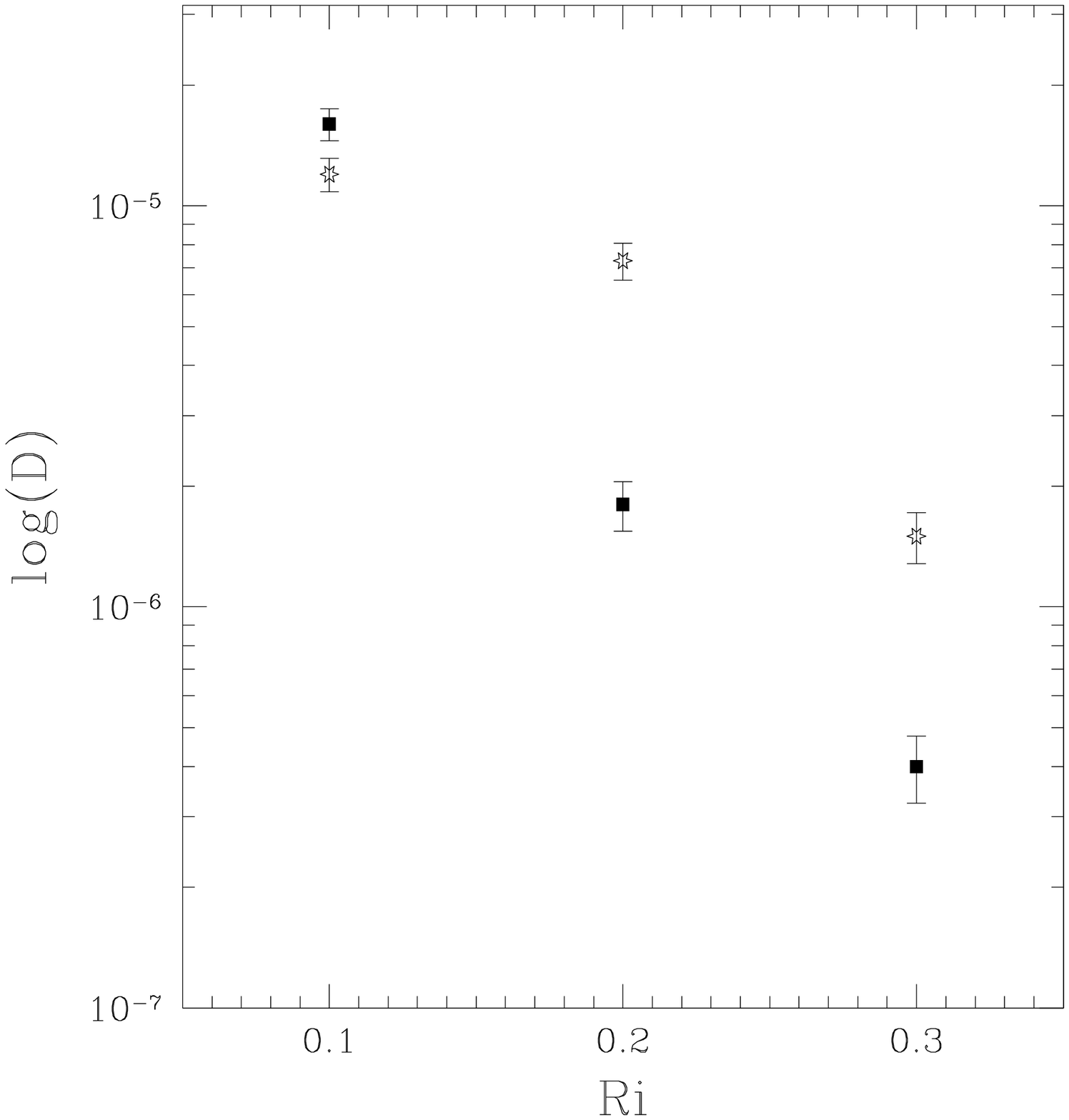}
\caption{Dimensionless diffusion coefficient as a function of
Richardson number for $B=0.001$ (stars) and $B=0.003$ (squares). 
}
\label{diff2}
\end{figure}

In Fig. \ref{diff2} we show the diffusion constant as a function of the
Richardson number for the different magnetic field strengths. The
errorbars indicate the residual scatter. Obviously, $D$ decreases with
increasing Richardson number similar to the non-magnetic case. But
furthermore, the power of the magnetic field to suppress the mixing
becomes evident. The diffusion coefficients for the cases with
$B=0.003$ are by a factor of a few smaller than the respective cases
with $B=0.001$ as long as Ri $> 0.1$. For Ri=0.1, however, the
diffusion coefficient for the strong magnetic field lies above the low
field case. \\

To make the Richardson number the only controlled parameter in our
study, constitutes a severe simplification of the factors that
determine the efficiency of mixing. In reality, the efficiency of
mixing will depend on the density stratification and the velocity
gradient separately, and not solely through the Richardson
number. Moreover, the mixing will depend on the exact shape of the
velocity profile and, only to a first approximation, on its first
derivative. Nevertheless, if one has to pick a single parameter to
describe the shear flow, e.g. for stellar evolution studies, the
Richardson number would be the most suitable one.

\subsection{3D simulations}

\begin{figure}
\plotone{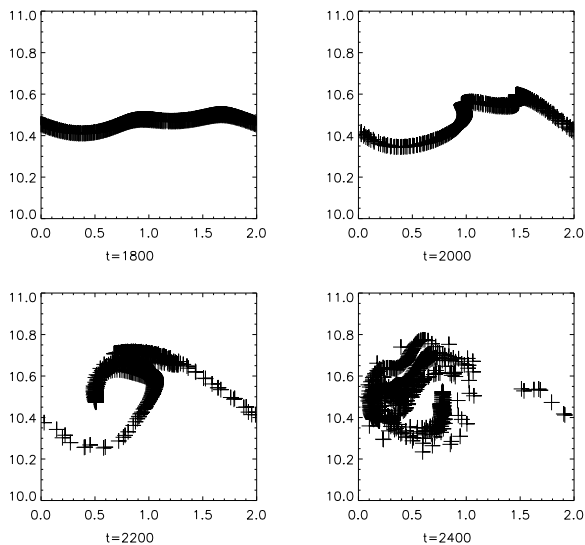}
\caption{Positions of tracer particles at various times for a 3D simulation with
Ri=0.2 and B=0.001. 
}
\label{tr3D_1}
\end{figure}

In the 3D simulations for the case with Ri=0.1 and $B=0.001$, we found
that the suppression of the primary vortex is less effective than in
the 2D simulations.  In Fig. \ref{tr3D_1} we have plotted the positions of the
tracer particles for the same case as in Fig. \ref{tr2D_1}, but now taken from
the 3D simulations. In the 3D simulations, contrary to the 2D case, a
big primary vortex forms which vigorously mixes the tracer
particles. This vortex is transitory and decays approximately within
one turnover time, as already observed in the unmagnetised case. The
final distribution of the tracer particles looks more like the
unmagnetized case than the 'laminar' picture in Fig. \ref{tr2D_1}. In 3D the
magnetic field does not seem to be able to suppress the formation of
the vortex. In the simulations with higher Richardson numbers the 3D
simulations yielded results more similar to the 2D simulations. This
results agree with the findings of Ryu, Jones \& Frank (2000) who also
noted that in 3D the magnetic field is more disruptive than in
2D. However, for the case in which the magnetic field enhanced the
mixing, i.e. with $B=0.003$ and Ri=0.1, we found that the enhancement
is somewhat less than the 2D simulations suggested. \\

This observation is fundamentally different from the simulations of
the non-magnetic Kelvin-Helmholtz instability. In the non-magnetic
case the differences between the mixing rates of the 2D and 3D
simulations were not as strong. Now, with the inclusion of magnetic
fields, the picture is different, and this constitutes one of the main
results of this paper.\\

\begin{figure}
\plotthree{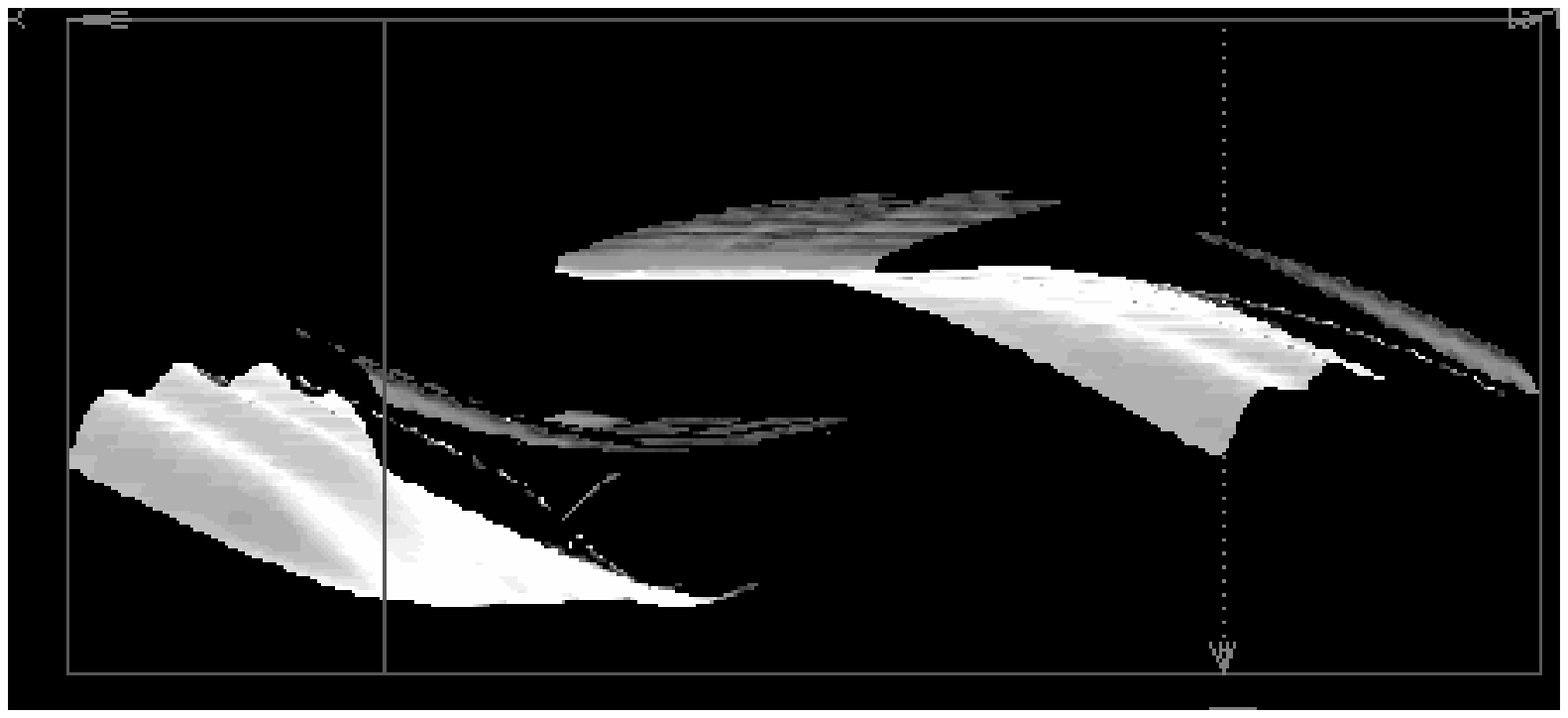}{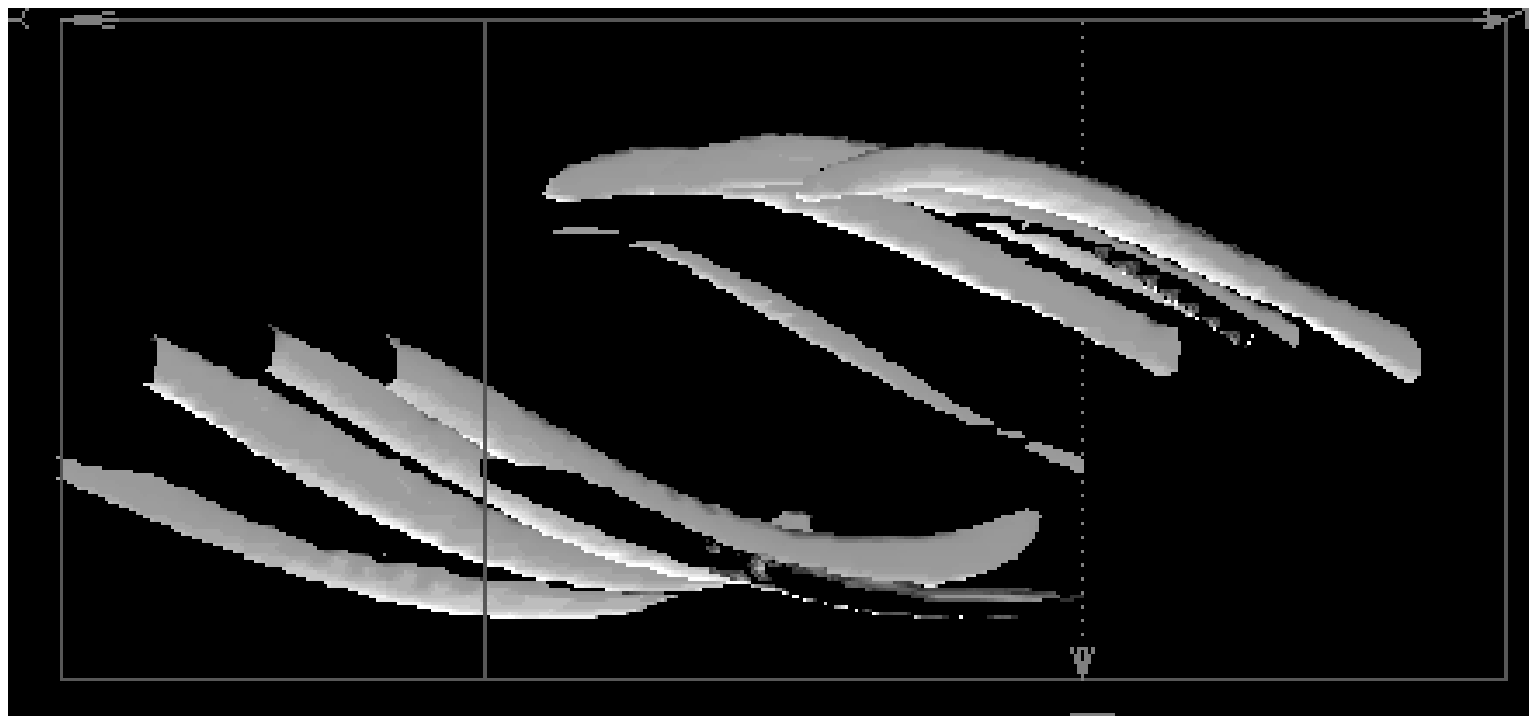}{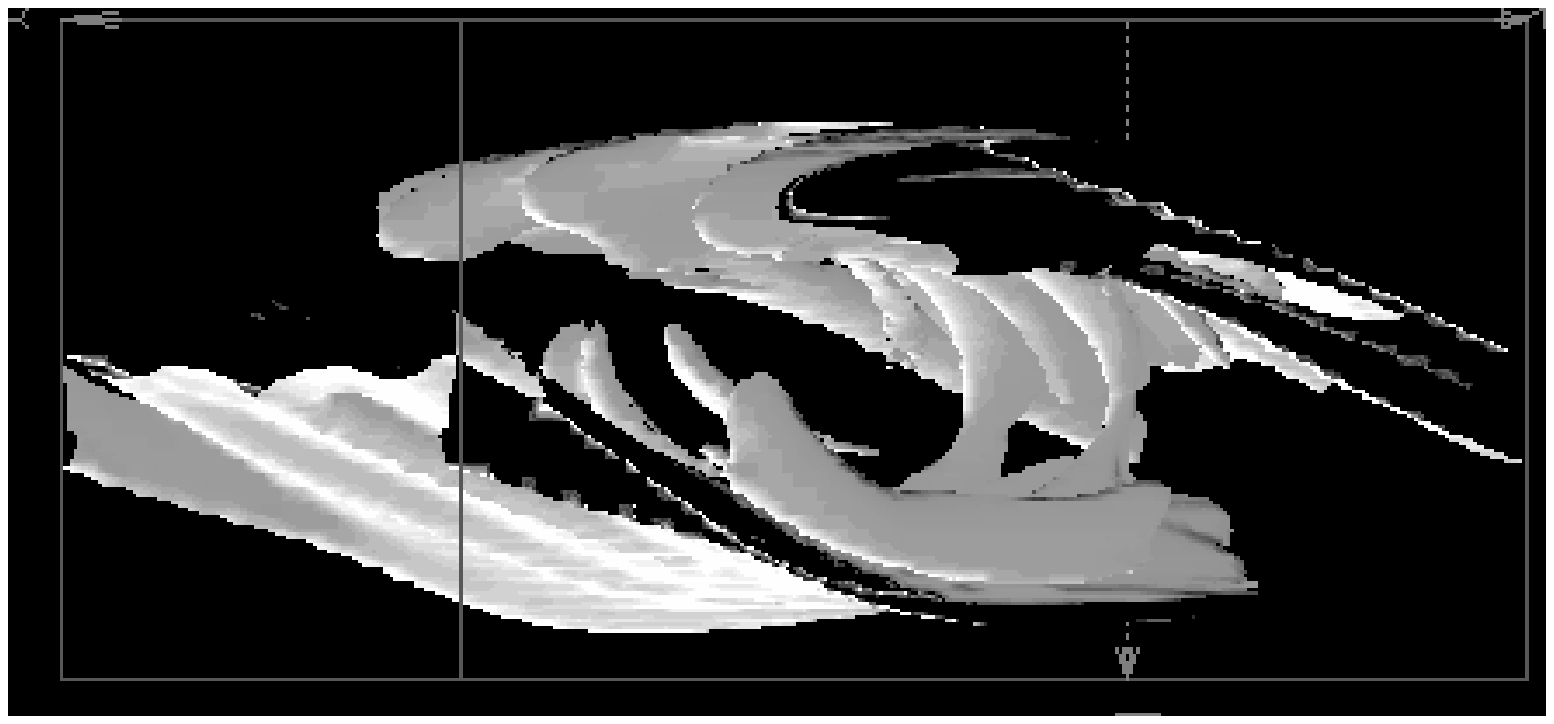}
\caption{Magnetic energy in a 3D simulation with Ri=0.1, $B=0.001$ at
times $t=2000$, $t=2300$ and $t=2500$.
}
\label{me3D_2}
\end{figure}

To investigate the origin of the discrepancy between the 2D and 3D
simulations, we visualised the magnetic energy in isosurface plots.
Figs. \ref{me3D_2} shows that the magnetic field becomes very
inhomogeneous and does not retain its symmetry in the
$y$-direction. One can observe that the magnetic field condenses into
6 - 8 flux tubes. In these flux tubes, the magnetic pressure is about
10 times as high as in the regions outside the flux tubes. This agrees
with the simulations by Ryu, Jones and Frank (2000) who observe that
the field becomes corrugated in the $y$-direction and then develops
into flux tubes. These flux tubes are aligned with the flow and have
the tendency to rise in the fluid due to their buoyancy.
Flux tubes of the same polarity repel each other which explains their
almost equidistant spacing in the horizontal which is seen in
Fig. \ref{me3D_2}. This repulsion grows with the strength of the
magnetic field. In our simulation we observed that the spacing between
the flux tubes is bigger for higher magnetic field strengths. In the
runs with a higher magnetic field fewer flux tubes formed. The lower
flux tubes are observed to rise in the fluid and during this rise
merge with neighbouring flux tubes. This transverse attraction of
rising flux tubes is a well known phenomenon that has been observed,
e.g., in the bipolar magnetic regions of the Sun. The attraction
between rising flux tubes is caused by purely hydrodynamical forces
(see, e.g. Parker 1979).\\

The flux tubes with their high field strengths become dynamically
important. They induce vortical motions in the $yz$-plane which
enhance the mixing rather than suppress it. In the run with $B=0.001$
and Ri=0.1 we found that the magnetic field is enhanced even more than
in the 2D simulations, namely by a factor of $\sim 15$ compared to
$\sim 12$ in 2D. For runs with higher Richardson numbers the
enhancement was similar or less than that found in the 2D
simulations.\\

\begin{figure}
\plotthree{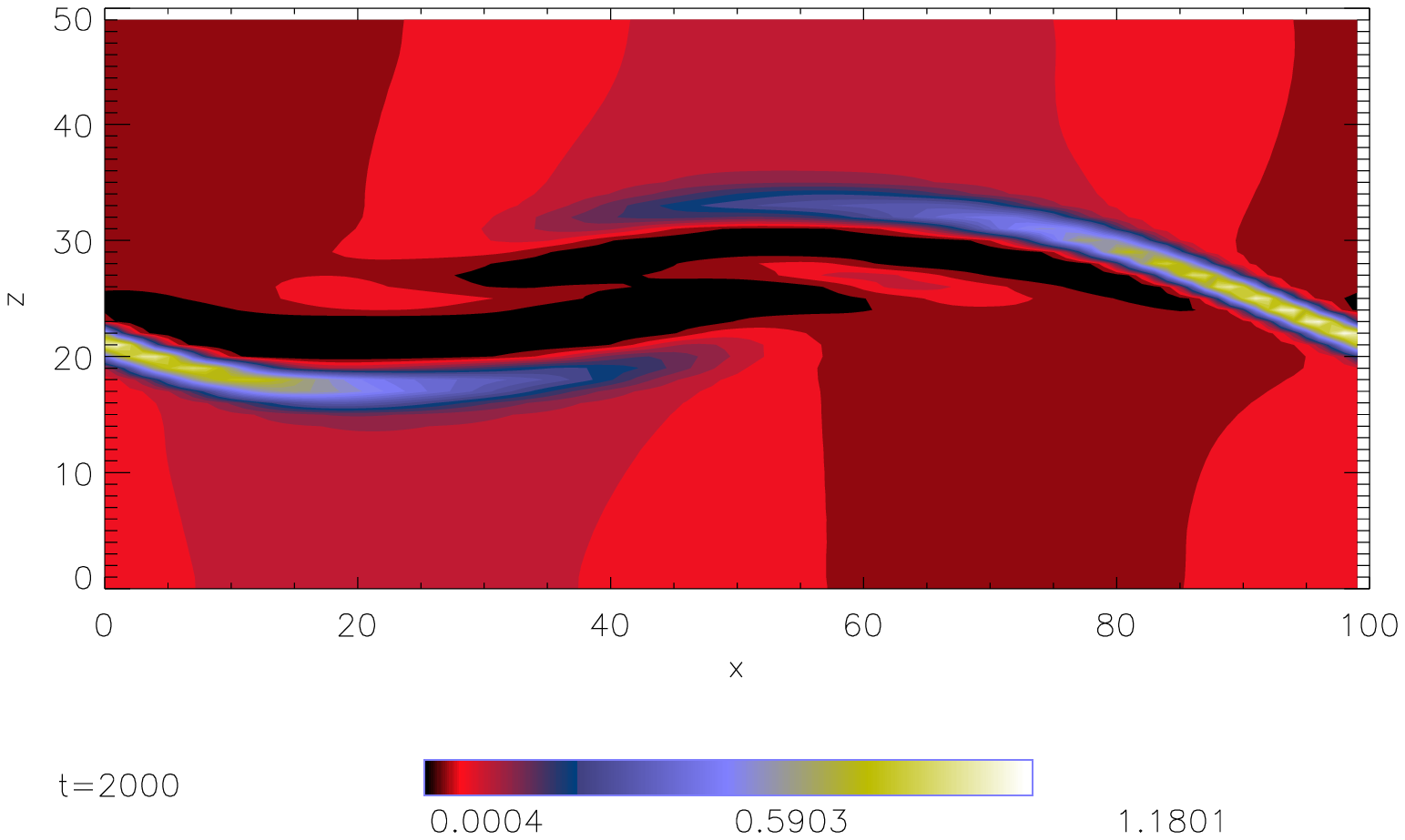}{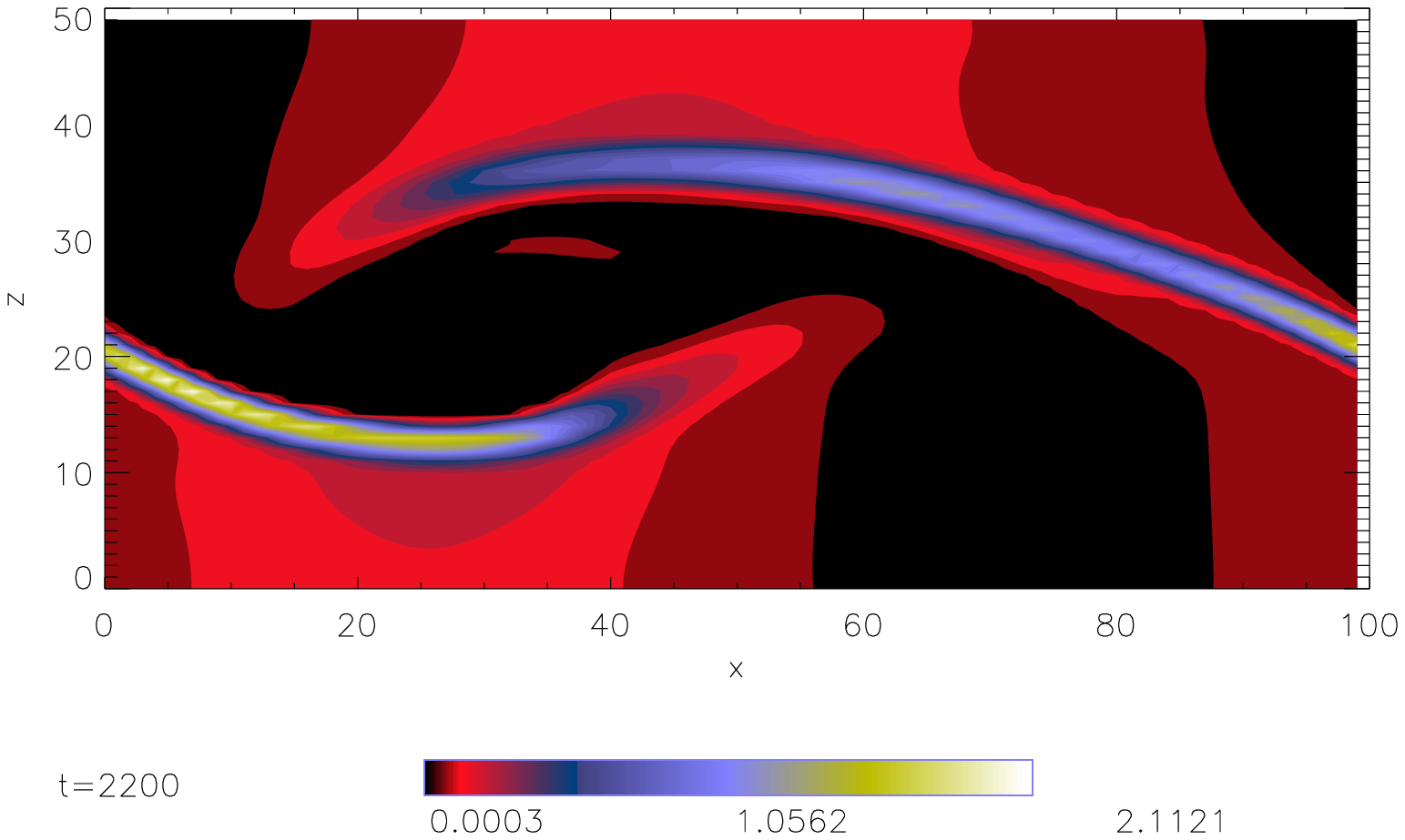}{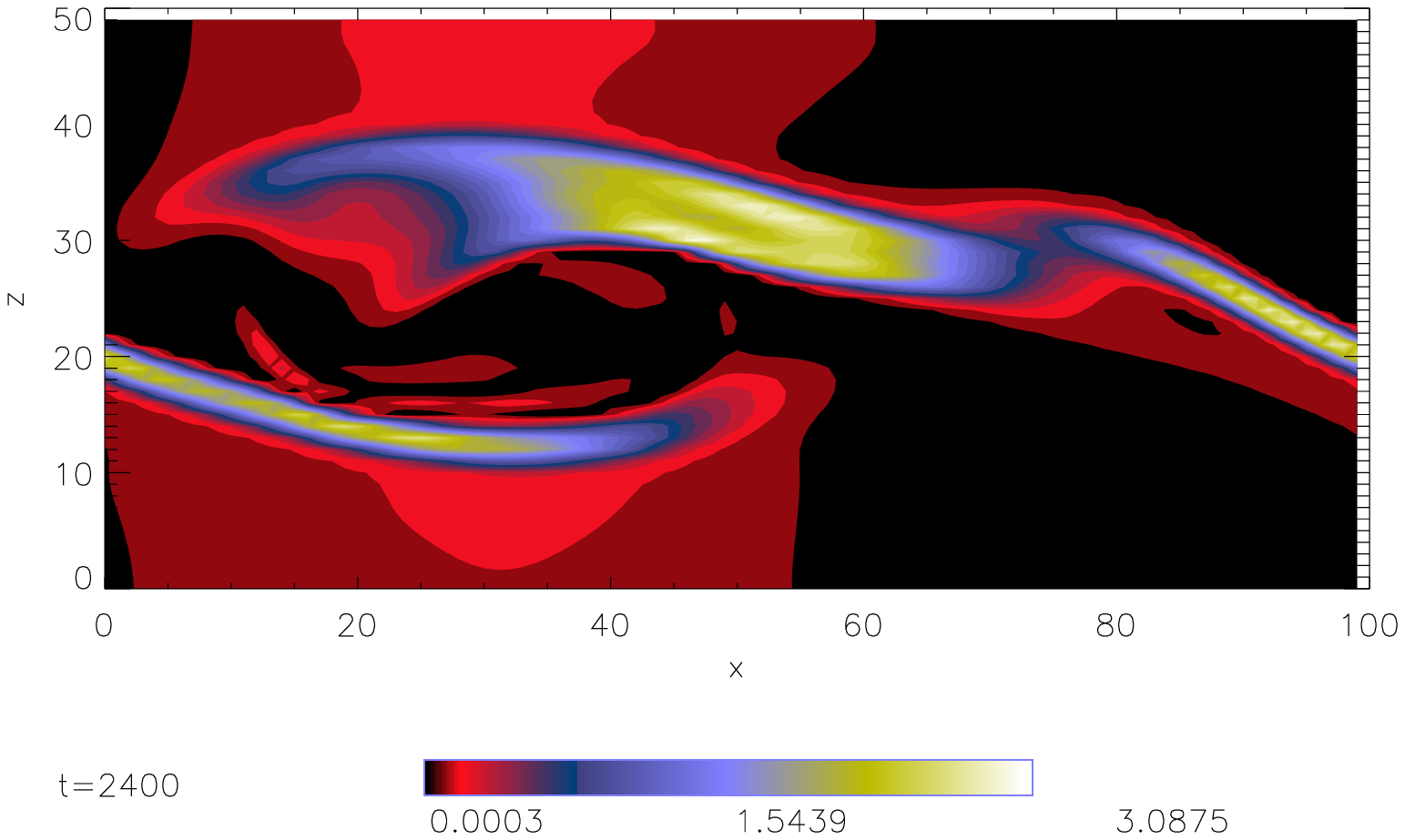}
\caption{Magnetic energy in
a slice along the $xz$-plane of a 3D simulation with Ri=0.2, $B=0.001$.
}
\label{me3D_1}
\end{figure}

From the positions of the tracer particles we, again, calculate a
diffusion coefficient. Its variation with the Richardson number is
shown in Fig. \ref{diff2}. In general, the diffusion coefficient is
higher than in the respective 2D case as discussed above. However, the
behaviour is similar: $D$ decreases with Ri and the magnetic field
suppresses the mixing for all cases except for Ri=0.1.\\

We tested the robustness of our results by changing the size of the
computational domain and the boundary conditions. We repeated some
simulations with periodic boundary conditions in the $y$- direction
and did not find any noteworthy differences to our results. The same
was true for simulations on a grid that was twice as large in the $z$-
direction. The nature of the boundaries changed the details of our
results but did not change the global morphology of the flow.

\subsection{Numerical viscosity}

One frequently voiced objection to these kinds of direct numerical
simulations is that Reynolds numbers as high as those encountered in
most astrophysical conditions are unattainable on current computers,
and that therefore the results are unrealistic. But, as pointed out by
Balbus, Hawley \& Stone (1996), this criticism is unjustified for
simulations of the shear instability. They argue, that in order to
simulate the onset of instability in a laminar flow, it is merely
necessary that the `typical' wavelength of the instability is resolved
by the numerical scheme and that the numerical diffusion at this
wavelength is less than the growth rate. This makes the simulation of
shear instabilities an easier task than the simulation of viscous
instabilities where in theory one would have to resolve everything
down to the viscous length scale.\\

Porter \& Woodward (1994) have estimated the Reynolds number of
hydrodynamical simulations based on a PPM (piecewise parabolic method)
code. The Reynolds number depends on the truncation error of the
finite-difference algorithm, the Courant number, the background
advection and the number of grid points. They found that the effective
Reynolds number is proportional to the third power of the number of
grid points, with the main dissipation occurring at short
wavelengths. Above this critical wavelength the diffusion was found to
be small. Since the ZEUS code uses piecewise linear functions instead
of piecewise parabolic functions, the truncation errors of ZEUS will
be larger than those of a PPM code. But even if they were only
proportional to the second power of the number of grid points in one
direction, we would still expect a numerical Reynolds number of around
$10^4$.\\

In ideal MHD simulations, numerical truncation errors also produce
some numerical resitivity. Rough estimates based on the work of Ryu et
al. (1995) suggest that the magnetic Reynolds number for our grid
is $< 100$.\\

If the flow is unbounded, there is no viscous boundary layer which
might interfere with the results. The nonlinear instabilities are
fundamentally inviscid in character. Therefore, for our purpose, we
only need a resolution capable of resolving a range of wavelengths for
which the numerical diffusion errors are less than the growth rates.
Finally, we should emphasize that we are not claiming to have
simulated magnetohydrodynamic turbulence.  We merely simulated
the Kelvin-Helmholtz instability and not full 3D turbulence which is a
much more difficult problem and which requires a much finer (and yet
unattained) numerical resolution.

\section{Conclusions}

In this paper we presented 2D and 3D simulations of the
Kelvin-Helmholtz instability in magnetised and stratified shear flows.

We found that a uniform magnetic field parallel to the shear flow
affects the evolution of the flow in a number of ways. In most cases
it suppresses the mixing as expected from linear stability
analyses. However, for flows with low Richardson numbers and strong
magnetic fields the evolution of the flow is strongly nonlinear and
the magnetic field now enhances the mixing in the shear layer. The
evolution of the magnetic shear instability is complex and rich:
magnetic field lines stretch, bend and reconnect. Thereby, the
magnetic field is locally enhanced. It forms flux tubes which, in
turn, influence the dynamics of the flow as they merge and rise
through the fluid. \\

For the cases in which the magnetic field suppresses shear
 instabilities, the 3D simulations reveal that a parallel magnetic
 field is less effective in suppressing the mixing than linear
 stability analyses and 2D simulations suggest. On the other hand, the
 field is less capable of enhancing the mixing for low Richardson
 numbers and strong magnetic fields as observed in 2D. The magnetic
 field concentrates in flux tubes which are dynamically important and
 enhance the mixing in the shear layer.  This leads to motions in the
 plane transverse to the flow which further destabilise the flow. We
 investigated the mixing rate for a number of initial conditions that
 were characterised by the Richardson number and the magnetic field
 strength. Using tracer particles we quantified the mixing by a
 heuristic diffusion coefficient which is plotted in Fig. \ref{diff2}. \\

However, a direct application of our results, especially
Fig. \ref{diff2}, should be treated with some caution. Our setup is
rather artificial and was primarily intended to give some insights
into the physics of the shear instability in magnetized flows. We have
performed a parameter study on a rather restricted set of parameters,
mostly because these simulations are computationally
expensive. Further studies will have to extend, both, the number as
well as the ranges of these parameters. The consequences of newly
found diffusion coefficients on stellar evolution and elemental
abundances, for example, are difficult to foresee. In stars, the speed
as well as the depth of mixing determine the balance between mixing
and nuclear burning. The effects of mixing will have to be studied in
detail as they depend sensitively on the conditions prevailing in the
star.\\

Finally, we should mention that a number of factors can inhibit or
facilitate mixing such as gradients in the chemical potential of the
fluid, diffusion of radiation and effects pertaining to the spherical
geometry.  In late-type stars strong chemical composition gradients
exist, which will have a stabilising effect on the
stratification. Therefore, especially for stellar evolution studies,
the chemical composition gradient is an important parameter, which
will have to be included in future work.

\section*{Acknowledgements}

A part of the simulations were performed on computers of the
Rechenzentrum Garching. We thank Phil Armitage and Henk Spruit for
helpful discussions.

\end{document}